\documentclass[twocolumn,secnumarabic,amssymb, nobibnotes, aps, prd]{revtex4-2}

\usepackage{graphicx}
\usepackage{dcolumn}
\usepackage{bm}
\usepackage[colorlinks=true,allcolors=blue]{hyperref}
\usepackage{float}
\usepackage{physics}
\usepackage{slashed}
\usepackage{relsize}
\usepackage{amsmath}
\usepackage{hyphenat}

\begin{document}
	
\title{Exploring Nucleon Structure and the Proton Mass Problem through Holographic QCD}
	
\author{Jiali Deng}
\affiliation{Institute of Particle Physics and Key Laboratory of Quark and Lepton Physics (MOS),
 Central China Normal University, Wuhan 430079, China}
\email{djl2022010355@mails.ccnu.edu.cn,
houdf@mail.ccnu.edu.cn
}	
\author{Defu Hou}
\affiliation{Institute of Particle Physics and Key Laboratory of Quark and Lepton Physics (MOS),
 Central China Normal University, Wuhan 430079, China}

\date{\today}
\begin{abstract}

Understanding the internal structure of the proton—including the distributions of quarks and gluons and their contributions to proton properties such as mass—remains a central challenge in quantum chromodynamics (QCD). While quark generalized parton distributions (GPDs) have been studied extensively, a consistent approach that simultaneously extracts quark parton distribution functions (PDFs), gravitational form factors (GFFs), and gluon GPDs from experimental constraints is still lacking. Moreover, the role of gluons in proton mass generation, particularly through the trace anomaly mechanism, requires deeper theoretical and phenomenological exploration. In this study, we begin by extracting quark GPDs in protons using a parameterization method based on the electromagnetic form factors provided by Light-Front Holographic QCD (LFHQCD), from which we derive both quark PDFs and their GFFs. Our contribution lies in adopting a distinct functional form for $w(x,t)$ and extending this approach to model gluon GPDs. Our calculations show consistency with experimental data and lattice QCD results and successfully reproduce soft Pomeron behavior. Furthermore, we investigate near-threshold $J/\psi$ production using gauge/string duality to quantify the contribution of the trace anomaly to the proton mass.
Our results indicate that the parameterization method provides a consistent framework for describing both quark and gluon structure, bridging GPDs, PDFs, and GFFs. The analysis of $J/\psi$ production confirms that the trace anomaly contributes significantly ($\sim 24\%$)  to the proton mass, with the calculated cross-section dependence on momentum transfer $t$ in agreement with experimental observations.  This work advances the understanding of proton structure by integrating quark and gluon degrees of freedom and elucidating the origin of proton mass within QCD.

\end{abstract}
	
\maketitle	

\section{Introduction}\label{sec:01_intro}

A central challenge in modern strong interaction physics is to unravel the origin of mass in the visible universe. While the Higgs mechanism endows elementary particles with their intrinsic masses, it accounts for only about $1\%$ of the proton's mass. The remaining $99\%$ emerges dynamically from the complex interactions of quantum chromodynamics (QCD), the theory of the strong force. Understanding this profound generation of mass from nearly massless constituents (quarks and gluons) represents one of the deepest questions in contemporary nuclear and particle physics \cite{Burkert:2023wzr}. 

This pursuit necessitates a detailed, three-dimensional picture of the proton's internal structure, a goal that has driven decades of experimental and theoretical effort. Generalized Parton Distributions (GPDs) have emerged as the fundamental theoretical framework for such a comprehensive description \cite{Guidal:2013rya}. These non-perturbative objects, functions of the longitudinal momentum fraction $x$, the skewness $\xi$, and the four-momentum transfer squared $t$, provide a unified picture of the nucleon. In specific limits, GPDs reduce to the well-known parton distribution functions (PDFs) and electromagnetic form factors (EFFs). Crucially, their first moments are linked to the gravitational form factors (GFFs), which encode the spatial distributions of mass, spin, and mechanical properties like pressure and shear forces inside the proton \cite{Pacetti:2014jai,Deng:2025azp,Nair:2024fit,Burkert:2023wzr,Deng:2025fpq}. Through a Fourier transform in transverse momentum transfer, GPDs offer access to the three-dimensional tomography of the nucleon, mapping partons in both longitudinal momentum and transverse position \cite{Guidal:2013rya, Diehl:2015uka}.

A pivotal decomposition of the proton mass within QCD separates it into four gauge-invariant contributions: the quark kinetic energy ($M_q$), the gluon kinetic energy ($M_g$), the quark mass term ($M_m$), and the trace anomaly term ($M_a$) \cite{Roberts:2016vyn,Lorce:2017xzd, Wang:2019mza,He:2021bof, Lorce:2021xku}. The trace anomaly, arising from the breaking of classical scale symmetry in quantum chromodynamics, is of particular contemporary interest. It is intrinsically connected to the gluonic structure of the proton and is believed to contribute significantly to its mass. A promising avenue to probe this elusive component is through the exclusive photoproduction of heavy quarkonia, such as $J/\psi$, near the production threshold in electron-proton scattering ($ep \rightarrow e'p'J/\psi$) \cite{Kharzeev:1998bz,ZEUS:2002wfj, Brodsky:2000zc, Frankfurt:2002ka,Gryniuk:2016mpk, Hatta:2018ina}. The dominance of the gluon-exchange mechanism in this process provides a direct window into the proton's gluonic GPDs and, by extension, the gravitational form factors related to the mass structure.

Despite significant advances from lattice QCD calculations and global fits to experimental data, a first-principles, model-consistent extraction of both quark and gluon GPDs and GFFs from a unified QCD-inspired framework remains a challenge. Furthermore, while the connection between threshold $J/\psi$ production and the trace anomaly has been theorized, a direct calculation of the cross section within a holographic QCD approach that naturally incorporates non-perturbative dynamics and explicitly uses computed GFFs as input is lacking. Most analyses rely on factorization theorems that may have limitations near threshold, and few studies self-consistently propagate the QCD scale evolution of coupling constants and anomalous dimensions into the final observable.

The light-front holographic QCD (LFHQCD) framework employed in this work is a phenomenological model, not a first-principles calculation from QCD. The results presented here are therefore inherently model-dependent. Nevertheless, LFHQCD provides a well-motivated and analytically tractable approach that captures key non-perturbative features of QCD, such as confinement and chiral symmetry breaking. We compare our predictions with experimental data and lattice QCD results to assess their reliability. We begin by deriving the proton's electromagnetic form factors within LFHQCD and subsequently construct the associated quark GPDs. From these, we self-consistently extract both the PDFs and the quark GFFs. We then generalize this formalism to determine the gluon GPDs and GFFs within the same consistent model. Our results are consistent with available experimental data and lattice QCD computations, while also exhibiting features characteristic of the soft Pomeron exchange. Finally, leveraging the gauge/string duality, we directly compute the cross section for exclusive $J/\psi$ photoproduction in $ep$ collisions. This holographic approach circumvents traditional factorization complexities by calculating the scattering amplitude in the gravitational dual theory. Crucially, we use our previously obtained GFFs as direct input and incorporate the energy-scale dependence via the two-loop QCD $\beta$-function and the anomalous dimension $\gamma_m$. This allows us to establish a direct, quantitative link between the measured cross section and the proton's mass decomposition. The primary objectives of this study are therefore: (1) to derive a unified set of quark and gluon GPDs and GFFs from the LFHQCD framework; (2) to compute the cross section for threshold $J/\psi$ photoproduction using these GFFs within a gauge/gravity duality model; and (3) to quantitatively extract the contribution of the QCD trace anomaly to the proton mass from the comparison of our calculation with experimental data. Our methodology combines non-perturbative holographic modeling with the constraints of QCD renormalization group flow.

The significance of this work is twofold. Theoretically, it demonstrates the power of holographic QCD to provide a consistent, parameter-lean description of multiple facets of nucleon structure—from electromagnetic form factors to gravitational properties. Practically, it provides a novel estimate of the trace anomaly’s contribution to the proton mass within the LFHQCD framework, offering a benchmark for future experiments at facilities like the Electron-Ion Collider (EIC). Our finding that the trace anomaly accounts for approximately 24.0\% of the proton mass provides further support for the understanding of the emergence of hadronic mass from QCD, consistent with existing lattice and phenomenological determinations.

In contrast to our previous studies, where Ref. \cite{Deng:2025fpq} utilized the VQCD model with distinct parameter sets for different observables and Ref. \cite{Deng:2025azp} employed a unified parameter set but omitted the trace anomaly in $J/\psi$ production, the present work advances these studies by extracting both quark and gluon PDFs to determine the gravitational form factors and including the trace anomaly contribution in the $J/\psi$ cross-section calculation.

The remainder of this paper is structured as follows. Section II details the calculation of quark and gluon GPDs within the proton using the LFHQCD approach. Section III presents the computation of the $J/\psi$ photoproduction cross section via gauge/string duality and our subsequent determination of the trace anomaly contribution. Finally, Section IV summarizes our key results and discusses their broader implications for strong interaction physics.

\section{Generalized parton distributions}\label{sec:02}

Among the various observables, the EFFs are the most straightforward to measure experimentally and compute theoretically, with a wealth of existing research \cite{A1:2010nsl,A1:2013fsc,Ye:2017gyb,Christy:2021snt,Park:2021ypf,Djukanovic:2023jag,Brodsky:2014yha,Sufian:2016hwn}. Therefore, in this section, we derive the proton's GPDs, PDFs, and GFFs starting from the EFFs obtained within the LFHQCD model.

\vspace{1em}
\noindent\textbf{The electromagnetic form factors}

\vspace{1em}
Proton EFFs are key observables for probing nucleon structure and dynamics. Elastic electron scattering off spin-1/2 protons is fully described by two form factors: the Dirac form factor $F_1$ and the Pauli form factor $F_2$ \cite{Brodsky:2014yha, Sufian:2016hwn}
​\begin{equation}
	\label{eq1}
\langle P'|J^{\mu}(0)|P\rangle=\bar{u}(P')[\gamma^{\mu}F_1(t)+\frac{i\sigma^{\mu\nu}q_\nu}{2M}F_2(t)]u(P),
\end{equation}
where $P$ and $P'$ are the initial and final four-momenta of the proton, respectively, $q=P'-P$ is the four-momentum transfer and $t=q^2=-Q^2$, $M$ is the proton mass, and \(\sigma^{\mu\nu} = \frac{i}{2}[\gamma^\mu, \gamma^\nu]\) is the commutator of the gamma matrices. In the light-front formalism, the Dirac and Pauli form factors can be extracted from the light-front spin-conserving and spin-flip matrix elements of the $J^{+}$ current:
​\begin{equation}
	\label{eq2}
\langle P',\uparrow|\frac{J^{+}(0)}{2P^+}|P,\uparrow \rangle=F_1(t),
\end{equation}
​\begin{equation}
	\label{eq3}
\langle P',\uparrow|\frac{J^{+}(0)}{2P^+}|P,\downarrow \rangle=\frac{q^1-iq^2}{2M}F_2(t),
\end{equation}

In the higher-dimensional gravity theory in the bulk, the spin-non-flip amplitude for the electromagnetic transition arises from the coupling between an external electromagnetic field $A_N(x,z)$ propagating in anti-de Sitter (AdS) space and a fermionic mode $\Psi_P(x,z)$, as expressed by the left-hand side of the equation below.
​\begin{equation}
 \begin{split}
	\label{eq4}
\int d^{4}xdz&\sqrt{g}\bar{\Psi}_P'(x,z)e^B_N\Gamma_BA^N(x,z)\Psi_P(x,z)\\
&\sim (2\pi)^4\delta^4(P'-P-q)\epsilon_\mu u(P')\gamma^\mu F_1(t)u(P),
 \end{split}
\end{equation}
where $\Gamma_B=(\gamma_\mu,-i\gamma_5)$. $e^B_N$ denotes the five-dimensional vielbein, with $N$ being the curved spacetime index and $B$ the local Lorentz index. Protons are represented by wave functions $\Psi_+$ and $\Psi_{-}$, which correspond to the nucleon's positive and negative chirality components, respectively.
​\begin{equation}
	\label{eq5}
\Psi_+(z)\sim z^{\tau+1/2}e^{-k^2z^2/2},\ \ \Psi_-(z)\sim z^{\tau+3/2}e^{-k^2z^2/2},
\end{equation}
where $\tau$ represents the twist of the component. These correspond, respectively, to a positive-chirality component with orbital angular momentum $L=0$ and a negative-chirality component with $L=1$, both sharing the same normalization. The spin-non-flip nucleon elastic form factor $F_1$ is expressed in terms of $\Psi_+$ and $\Psi_-$
\begin{equation}
	\label{eq6}
F_1(t)=\Sigma g_{\pm}\int\frac{dz}{z^4}V(t,z)\Psi_{\pm}^2(z).
\end{equation}
The effective charges $g_{\pm}$ are fixed by the underlying spin-flavor structure and $V(t,z)$ represents the five-dimensional mode of the electromagnetic field

The nucleon spin-flip electromagnetic form factor arises from the non-minimal coupling term
\begin{equation}
 \begin{split}
	\label{eq7}
&\int d^{4}xdz\sqrt{g}\bar{\Psi}_P'(x,z)e^M_Ae^N_B[\Gamma^A,\Gamma^B]F_{MN}(x,z)\Psi_P(x,z)\\
&\sim (2\pi)^4\delta^4(P'-P-q)\epsilon_\mu u(P')\frac{\sigma^{\mu\nu}q_\nu}{2M} F_2(t)u(P),
 \end{split}
\end{equation}
where $[\Gamma^A,\Gamma^B]=2\eta^{AB}$ and $F_{MN}=\partial_MA_N-\partial_NA_M$ represents the electromagnetic field strength tensor.
The spin-flip nucleon elastic form factor $F_2$ is
\begin{equation}
	\label{eq8}
F_2(t)=\chi \int\frac{dz}{z^3}V(t,z)\Psi_{+}(z)\Psi_{-}(z),
\end{equation}
where $\chi$ is determined by the measured magnetic moment from experiments.

In light-front quantization, a hadron state $|H\rangle$ is expressed as a superposition of infinitely many Fock components $|N\rangle$, $|H\rangle=\sum_N\psi_{N/H}|N\rangle$,where $\psi_{N/H}$ denotes the light-front wave function(LFWF) of the N-particle component with normalization $\sum_N|\psi_{N/H}|^2=1$. Thus, the form factor is expressed as an infinite sum of contributions
\begin{equation}
	\label{eq9}
F_H(t)=\Sigma_\tau \varsigma_\tau F_\tau(t),
\end{equation}
where normalization at $t=0, F_\tau(0)=1, F_H(0)=1$ indicates that $\Sigma_\tau \varsigma_\tau=1$.

\vspace{1em}
\noindent\textbf{Quark GPDs}

\vspace{1em}
In LFHQCD, the electromagnetic form factors of protons can be expressed as \cite{Brodsky:2014yha, Sufian:2016hwn}
\begin{equation}
	\label{eq1}
F_{1}^{p}(t)=F_{\tau=3}(t),
\end{equation}
\begin{equation}
	\label{eq2}
F_{2}^{p}(t)=\chi_{p}[(1-\gamma_{p})F_{\tau=4}(t)+\gamma_{p}F_{\tau=6}(t)],
\end{equation}
where $\tau$ is the twist of the Fock state: $\tau = 3$ corresponds to the three-quark state with $L=0$, $\tau = 4$ to states with $L=1$, and $\tau = 6$ to higher Fock states. $\chi_{p}$ and $\gamma_{p}$ represent the proton anomalous magnetic moment and the probability of higher Fock state, respectively. The electromagnetic form factors of neutrons can be expressed as
\begin{equation}
	\label{eq3}
F_{1}^{n}(t)=-\frac{r}{3}[F_{\tau=3}(t)-F_{\tau=4}(t)],
\end{equation}
\begin{equation}
	\label{eq4}
F_{2}^{n}(t)=\chi_{n}[(1-\gamma_{n})F_{\tau=4}(t)+\gamma_{n}F_{\tau=6}(t)],
\end{equation}
where $r$, $\chi_{n}$ and $\gamma_{n}$ denote the $SU(6)$ breaking parameter, neutron anomalous magnetic moment, and probability of higher Fock state, respectively. $F_{1}^{N}(t)$ and $F_{2}^{N}(t)$ represent the Dirac form factor and Pauli form factor of the nucleon, respectively, $N=p,n$. The form factor of any twist can be expressed as \cite{deTeramond:2018ecg}
\begin{equation}
	\label{eq5}
F_{\tau}(t)=\frac{1}{N(\tau)}B(\tau-1,\frac{1}{2}-\frac{t}{4\lambda}),
\end{equation}
where $N(\tau)=\Gamma(\tau-1)\Gamma(1/2)/\Gamma(\tau-1/2)$ is the renormalization constant and $B(u,\nu)=\Gamma(u)\Gamma(\nu)/\Gamma(u+\nu)$ is the Euler beta function. For integer $\tau$, Eq.(\ref{eq5}) generates the pole structure
\begin{equation}
	\label{eq6}
F_{\tau}(t)=\frac{1}{(1-\frac{t}{M_{0}^{2}})(1-\frac{t}{M_{1}^{2}})(1-\frac{t}{M_{\tau-2}^{2}})}.
\end{equation}
where $M_{n}^{2}=4\lambda(n+\frac{1}{2})$, $n=0,1,2... \tau-2$. Eq.(\ref{eq5}) can also be written as $B(\tau-1,1-\alpha(t)$ with Regge trajectory
\begin{equation}
	\label{eq7}
\alpha(t)=\frac{1}{2}+\frac{t}{4\lambda}.
\end{equation}
This is the trajectory of the vector meson $\rho$ and $\lambda$ is fixed by the ground state mass of the $\rho$ meson: $\lambda=(0.548 \mathrm{GeV})^2$.

The integral form of the form factor is
\begin{equation}
	\label{eq8}
F_{\tau}(t)=\frac{1}{N(\tau)}\int_{0}^{1}\frac{dw(x,t)}{dx}w(x,t)^{\frac{-t}{4\lambda}-\frac{1}{2}}[1-w(x,t)]^{\tau-2}dx,
\end{equation}
where $x$ represents the longitudinal momentum fraction with $x\in (0,1)$ and $w$ satisfies the constraint conditions $w(0,t)=0$ and $w(1,t)=1$. The form factor remains invariant under any choice of $w(x,t)$ that satisfies the conditions.

The electromagnetic form factor can be written as the zeroth order moment of GPD: $F_{\tau}(t)=\int_{0}^{1}H_{\tau}(x,t)dx$, and then we can obtain
\begin{equation}
	\label{eq10}
H_{\tau}(x,t)=q_{\tau}(x,t)e^{tf(x,t)},
\end{equation}
with 
\begin{equation}
	\label{eq11}
q_{\tau}(x,t)=\frac{1}{N(\tau)}\frac{dw(x,t)}{dx}[1-w(x,t)]^{\tau-2}w(x,t)^{-1/2},
\end{equation}
\begin{equation}
	\label{eq11}
f(x,t)=-\frac{1}{4\lambda}\mathrm{ln}(w(x,t)),
\end{equation}
where $q_{\tau}(x,0)$ and $f(x,t)$ represent the PDF and the profile function, respectively.

For $x\rightarrow 0$, GPD can be expressed as \cite{Goeke:2001tz}
\begin{equation}
	\label{eq12}
H_{\tau}(x,t)\sim x^{-t/4\lambda}q_{\tau}(x,0),
\end{equation}
 which motivated by Regge theory, describes small-$x$ behavior. For $x\rightarrow 1$, GPD can be expressed as \cite{Brodsky:1979qm, Blankenbecler:1974tm,Lyubovitskij:2020otz}
\begin{equation}
	\label{eq13}
q_{\tau}(x,t)\sim (1-x)^{2\tau-3},
\end{equation}
which corresponds exactly to the Drell-Yan inclusive counting rule for large $x$.

Combining equations Eq.(\ref{eq12}) and Eq.(\ref{eq13}), we assume that the form of $w(x,t)$ is
\begin{equation}
	\label{eq14}
w(x,t)=x^{1-x}e^{-c_0(1-2x)(1-x)^2}e^{c_1x^{0.5}(1-x)^3t/(1+c_3t^2)},
\end{equation}
where the parameters $c_0=0.1$, $c_1=4.2$, and $c_3=0.05$ are obtained by fitting the experimental PDF data for the u quark and the lattice gravitational form factors for the u quark. $c_0$ is fixed by matching the quark PDF $xu(x)$ extracted from our light-front holographic QCD model to the PDF4LHC15 data at $x = 0.001$ (the smallest $x$ value provided by the experimental dataset). Meanwhile, $c_1$ is determined by fitting the computed gravitational form factor $A_u(Q^2)$ to lattice QCD GFF data. $c_3=0.05$ is chosen such that the gravitational form factors remain unchanged in the small-$t$ region, but adequately suppress their contributions in the large-$t$ region. Within the scope of our research, we have also performed a sensitivity check by varying the coefficient $c_3$ by $\pm20\%$ and examining the resulting change in the form factor. The form factor changes by less than $2.5\%$ under this variation, indicating that our results are stable against the uncertainty in the coefficient. This parameter remains unchanged across different flavors.

The sum rules relating the GPDs and the electromagnetic form factors \cite{Vega:2010ns}. The PDF of quarks are
\begin{equation}
	\label{eq15}
u(x,0)=(2-\frac{r}{3})q_{\tau=3}(x,0)+\frac{r}{3}q_{\tau=4}(x,0),
\end{equation}
\begin{equation}
	\label{eq16}
d(x,0)=(1-\frac{2r}{3})q_{\tau=3}(x,0)+\frac{2r}{3}q_{\tau=4}(x,0),
\end{equation}
\begin{figure}
	\centering
	\includegraphics[width=8.5cm]{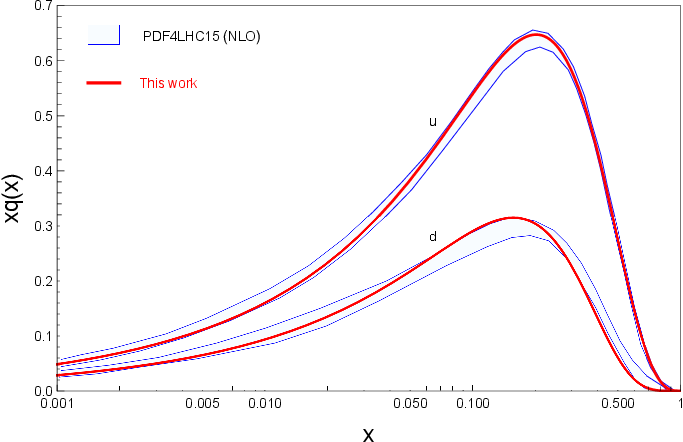}
	\caption{\label{figure}Comparison of the $xq(x)$ in the proton (red curve) with experimental data (blue band) \cite{Lin:2017snn} at $\mu^2=4\ \mathrm{GeV}^2$.}
\end{figure}

The gravitational form factors of valence quarks are
\begin{equation}
	\label{eq17}
A_u(t)=\int_{0}^{1}xu(x,t)e^{tf(x,t)},
\end{equation}
\begin{equation}
	\label{eq18}
A_d(t)=\int_{0}^{1}xd(x,t)e^{tf(x,t)},
\end{equation}
As shown in Fig.1 and Fig.2, our calculation results are consistent with experimental measurements and the lattice. 
\begin{figure}
	\centering
	\includegraphics[width=8.5cm]{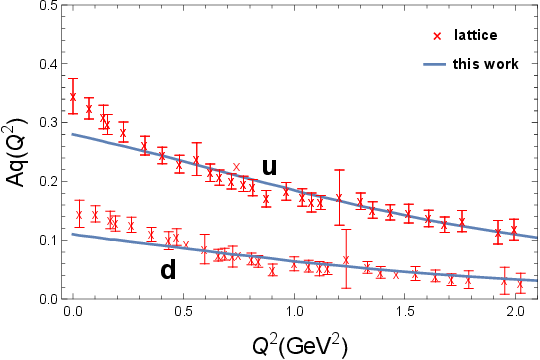}
	\caption{\label{figure}Comparison of the $u/d$ quark GFF $A_q(Q^2)$ in the proton (blue curve) with lattice QCD results (red crosses) \cite{Hackett:2023rif} at $\mu^2=4\ \mathrm{GeV}^2$.}
\end{figure}

\vspace{1em}

\noindent\textbf{Gluon GPDs}

\vspace{1em}
The gluon PDF in the proton can be written as the sum of all contributing Fock states, $g(x)=\sum_{\tau}\varsigma_{\tau}g_{\tau}(x)$. In this section, we only consider the contribution of the lowest state $\tau=4$.

Assuming the $g(x,t)$ and the profile $f(x,t)$ function of the gluon are
\begin{equation}
	\label{eq17}
g(x,t)=N\frac{dw(x,t)}{dx}[1-w(x,t)]^{2}w(x,t)^{c_2}, 
\end{equation}
\begin{equation}
	\label{eq17}
f(x)=-\frac{1}{4\lambda_{g}}\mathrm{ln}(w(x,t)),
\end{equation}
The form of $w(x,t)$ is the same as that of quarks, only the parameters are different:$c_0=-1.2,c_1=-2.3,N=3.8$. The parameter $N$ is determined by the gluon gravitational form factor $A_g(0)$ calculated from the lattice.
\begin{figure}
	\centering
	\includegraphics[width=8.5cm]{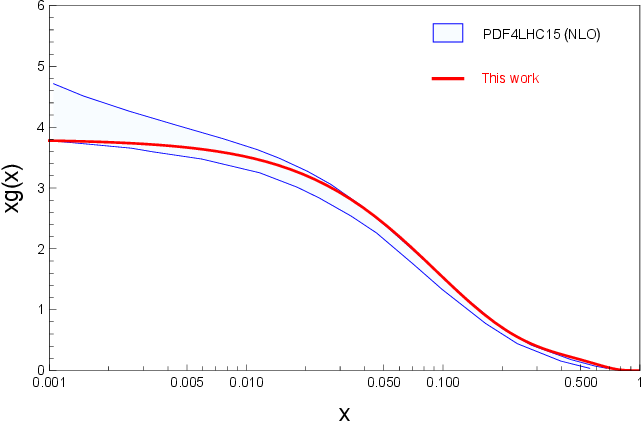}
\caption{\label{figure}Comparison of the gluon distribution $xg(x)$ in the proton (red curve) with experimental data (blue band) \cite{Lin:2017snn} at $\mu^2=4\ \mathrm{GeV}^2$.}
\end{figure}
\begin{figure}
	\centering
	\includegraphics[width=8.5cm]{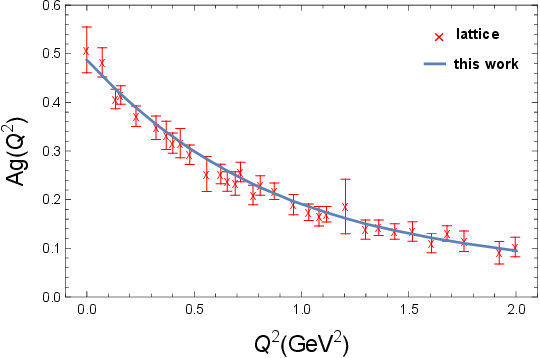}
	\caption{\label{figure}Comparison of the gluon GFF $Ag(Q^2)$ in the proton (blue curve) with lattice QCD results (red crosses) \cite{Hackett:2023rif} at $\mu^2=4\ \mathrm{GeV}^2$.}
\end{figure}

By analogy with equation Eq.(\ref{eq5}), we can obtain the mass spectrum as
\begin{equation}
	\label{eq18}
M_{n}^{2}=4\lambda_{g}(n+1+c_2),\ n=0,1,2.
\end{equation}

As noted in Ref.\cite{ParticleDataGroup:2022pth}, soft interactions are crucial in high-energy collisions. In particular, the Pomeron - the Regge trajectory governing diffractive processes - dominates small-angle, high-energy scattering, where perturbative QCD is no longer applicable. Since the inception of QCD, the Pomeron has been linked to the exchange of two or more gluons \cite{Gribov:1983ivg}. It interacts with hadrons as a rank-two tensor and exhibits strong coupling to gluons \cite{Ewerz:2013kda, Britzger:2019lvc}. Consequently, we employ the soft Pomeron model proposed by Donnachie and Landshoff \cite{Donnachie:1992ny}, incorporating their effective Regge trajectory.
\begin{equation}
	\label{eq19}
\alpha_{p}(t)=\alpha_{p}(0)+\alpha'_{p}t.
\end{equation}
where intercept $\alpha_{p}(0)=1.08$ and slope $\alpha'_{p}\simeq (0.20-0.25) \mathrm{GeV}^{-2}$, and it is interpreted as a two gluons bound state in QCD \cite{ParticleDataGroup:2022pth}. The Pomeron exchange mechanism corresponds to the graviton in the dual AdS theory \cite{Brower:2006ea, Costa:2012fw, Amorim:2021gat}.

Since the graviton is massless ($M_0=0\ \mathrm{GeV}$), we derive $c_2=-1$ from Eq.(\ref{eq18}), and then obtain the first excited state mass $M_1\simeq (2-2.24)\ \mathrm{GeV}$ with $\lambda_{g}=\frac{1}{4\alpha'_{p}}\simeq(1-1.14)\ \mathrm{GeV}^{2}$. Our result is relatively close to the predicted mass $M\simeq (1.86-2.4)\ \mathrm{GeV}$ of the $2^{++}$ glueball by other models \cite{Carlson:1982er, Morningstar:1999rf, Meyer:2004jc, Chen:2015zhh, Chen:2021bck}. In QCD, the ground state of gravitons does not participate in strong interactions. The first excited state of the graviton participates in strong interactions, manifested as a soft Pomeron.

Starting from the proton's electromagnetic form factor in the LFHQCD framework, we express it as an integral over the longitudinal momentum fraction $x$. By combining the asymptotic behaviors of GPDs at both small and large $x$, we derive the form of $\omega(x,t)$ and subsequently obtain the quark GPDs inside the proton. As illustrated in Fig.1, our calculated quark parton distribution functions demonstrate excellent agreement with experimental measurements. The quark gravitational form factors also show good overall consistency with lattice QCD results, except for some discrepancies at low momentum transfer in Fig.2. These differences stem from non-negligible contributions of sea quarks to the gravitational form factors at small $Q^2$, which rapidly decrease as $Q^2$ increases.

We assume the gluon GPDs share the same functional form as the quark GPDs, but with different parameters. As shown in Figs.3 and 4, our results demonstrate excellent agreement with both experimental measurements and lattice QCD calculations. The high-energy scattering processes probing gluon contributions are mediated through soft Pomeron exchange. In QCD, the Pomeron is interpreted as a $2^{++}$ bound state of two gluons, while in the dual AdS theory it corresponds to a graviton. By utilizing the masslessness of the graviton, we can fix parameter $c_2=-1$ and subsequently obtain the mass of its first excited state as $M_1=(2-2.24)\ \mathrm{GeV}$. This mass range agrees well with predictions for the $2^{++}$ glueball mass from lattice simulations and other theoretical models.

\section{the proton mass problem in Holographic QCD}\label{sec:03}

The Higgs boson generates the masses of quarks. However, its connection to the proton mass and consequently the mass of nuclei and atoms is a separate issue. The mass of quarks inside the proton, approximately 9 MeV, originates directly from the Higgs mechanism, while the proton's total mass is 938 MeV. A key question emerges: What is the origin of the proton mass? In this section, we employ holographic QCD to compute the $J/\Psi$ scattering cross section and investigate the contribution of the QCD trace anomaly to the proton mass.

\vspace{1em}

\noindent\textbf{Nucleon mass decomposition}

\vspace{1em}
In order to study the question of the origin of the proton mass, One should begin with the QCD energy-momentum tensor(EMT) \cite{Hatta:2018sqd,Hatta:2018ina,Lorce:2017xzd,Wang:2022tzw}. It is represented as
\begin{equation}
	\label{eq20}
T^{\mu\nu}=-F^{\mu\lambda}_{a}F^{a\nu}_{\lambda}+\frac{\eta^{\mu\nu}}{4}F^{\alpha\beta}_{a}F^{a}_{\alpha\beta}+i\bar{\Psi}\gamma^{(\mu}D^{\nu)}\Psi=T^{\mu\nu}_{g}+T^{\mu\nu}_{q},
\end{equation}
where $A^{(\mu}B^{\nu)}\equiv\frac{A^{\mu}B^{\nu}+B^{\mu}A^{\nu}}{2}$. $F^{\mu\lambda}_{a}$, $\eta^{\mu\nu}$, $\Psi$, $\gamma^{\mu}$ and $D^{\nu}$ are the gluon field strength, the Minkowski metric, the quark field, the four-dimensional gamma matrix and the covariant derivative, respectively. The trace of the EMT can be represented as
\begin{equation}
	\label{eq21}
T^{\alpha}_{\ \alpha}=\frac{\beta(g)}{2g}F^{\alpha\beta}_{a}F^{a}_{\alpha\beta}+m(1+\gamma_{m})\bar{\Psi}\Psi,
\end{equation}
where $\beta(g)$, $m$ and $\gamma_{m}$ represent the QCD beta function, the current quark mass and the anomalous dimension, respectively. The tensor can be decomposed into traceless and trace components
 \begin{equation}
	\label{eq22}
T^{\mu\nu}=(T^{\mu\nu}-\frac{\eta^{\mu\nu}}{4}T^{\alpha}_{\ \alpha})+\frac{\eta^{\mu\nu}}{4}T^{\alpha}_{\ \alpha}=T_{1}^{\mu\nu}+T_{2}^{\mu\nu}.
\end{equation}
The traceless component $T_{1}^{\mu\nu}$ can be further separated into gluon and quark parts $T_{1}^{\mu\nu}=T_{g,kin}^{\mu\nu}+T_{q,kin}^{\mu\nu}$ which represent the contribution of kinetic energy and the trace component $T_{2}^{\mu\nu}$ can also be written in two parts $T_{2}^{\mu\nu}=T_{m}^{\mu\nu}+T_{a}^{\mu\nu}$, namely the contribution of the current quark and anomaly. Then we can obtain
 \begin{equation}
	\label{eq23}
T^{\mu\nu}=T_{q,kin}^{\mu\nu}+T_{g,kin}^{\mu\nu}+T_{m}^{\mu\nu}+T_{a}^{\mu\nu}.
\end{equation}
The matrix elements of EMT can be represented as
 \begin{equation}
	\label{eq24}
\langle P|T^{\mu\nu}|P\rangle=2P^{\mu}P^{\nu},  \langle P|T^{\alpha}_{\ \alpha}|P\rangle=2M^{2}.
\end{equation}
Following Lorentz symmetry, one can parameterize their matrix elements as
 \begin{equation}
	\label{eq25}
\langle P|T_{q,kin}^{\mu\nu}|P\rangle=2a(P^{\mu}P^{\nu}-\frac{\eta^{\mu\nu}}{4}M^{2}),
\end{equation}
 \begin{equation}
	\label{eq26}
\langle P|T_{g,kin}^{\mu\nu}|P\rangle=2(1-a)(P^{\mu}P^{\nu}-\frac{\eta^{\mu\nu}}{4}M^{2}),
\end{equation}
 \begin{equation}
	\label{eq27}
\langle P|T_{m}^{\mu\nu}|P\rangle=\frac{1}{2}b\eta^{\mu\nu} M^{2},
\end{equation}
 \begin{equation}
	\label{eq28}
\langle P|T_{a}^{\mu\nu}|P\rangle=\frac{1}{2}(1-b)\eta^{\mu\nu} M^{2}.
\end{equation}
Then the proton mass can be expressed as
 \begin{equation}
	\label{eq29}
M=M_{q}+M_{g}+M_{m}+M_{a},
\end{equation}
with 
 \begin{equation}
 \begin{split}
	\label{eq30}
&M_{q}=\frac{3a}{4}M,\ \ \  M_{g}=\frac{3(1-a)}{4}M,\\  
&M_{m}=\frac{b}{4}M,\ \ \  M_{a}=\frac{1-b}{4}M.
\end{split}
\end{equation}
The parameter $a$ is determined by the twist-two operator matrix elements for both quarks and gluons and the parameter $b$ is related to the twist-four operator $F^2$.

\vspace{1em}

\noindent\textbf{Exclusive photoproduction of $J/\psi$ in $ep$ scattering}

\vspace{1em}
In this section, We begin with a concise overview of the $ep\rightarrow e'\gamma^{*}p\rightarrow e'p'J/\psi$ process kinematics. Our analysis will be conducted in the center-of-mass frame of the $\gamma^{*}-p$ system \cite{Mamo:2019mka}. Then we can obtain
 \begin{equation}
	\label{eq31}
|\vec{q}_{\gamma}|=\frac{1}{2\sqrt{s}}\sqrt{s^2-2(q_\gamma^2+M^{2})s+(q_\gamma^2-M^{2})^2},
\end{equation}
 \begin{equation}
	\label{eq32}
|\vec{q}_{V}|=\frac{1}{2\sqrt{s}}\sqrt{s^2-2(M_{V}^{2}+M^{2})s+(M_{V}^{2}-M^{2})^2},
\end{equation}
where $q_\gamma^2$ is the squared four-momentum of the virtual photon. $\vec{q}_{\gamma}$, $\vec{q}_{V}$, $M_{V}$ and $s=W^2=(p_{1}+q_{1})^2$ are the three momenta of the virtual photon, the three momenta of the $J/\psi$, the mass of the $J/\psi$ and the virtual photon-proton c.m. energy, respectively. The square of transferred momentum can be expressed as
 \begin{equation}
	\label{eq33}
t=q_\gamma^2+M_{V}^{2}-2E_{\gamma}E_{V}+2|q_{\gamma}||q_{V}|cos\theta,
\end{equation}
where $E_{\gamma}$, $E_{V}$ and $\theta$ represent the energy of the virtual photon, the energy of the $J/\psi$ and scattering angle, respectively.

The cross section of the photoproduction process takes the form
 \begin{equation}
 \begin{split}
	\label{eq34}
\sigma_{\gamma p\rightarrow p J/\psi}=\frac{e^2\int dt\langle P|\epsilon\cdot J(q)|P'k\rangle \langle P'k|\epsilon^{*}\cdot J(q)|P\rangle}{64\pi MKW|P_{cm}|},
\end{split}
\end{equation}
where $K=\frac{W^2-M^2}{2M}$ and $P_{cm}$ represents the incoming proton momentum in the c.m. frame
 \begin{equation}
	\label{eq35}
|P_{cm}|^2=\frac{W^4-2W^2(M^2+q_\gamma^2)+(M^2-q_\gamma^2)^2}{4W^2}.
\end{equation}
The t-transfer in the low-s region is constrained between $t_{min}\equiv t|_{cos \theta=1}$ and $t_{max}\equiv t|_{cos \theta=-1}$, as shown in Fig.5.
\begin{figure}
	\centering
	\includegraphics[width=8.5cm]{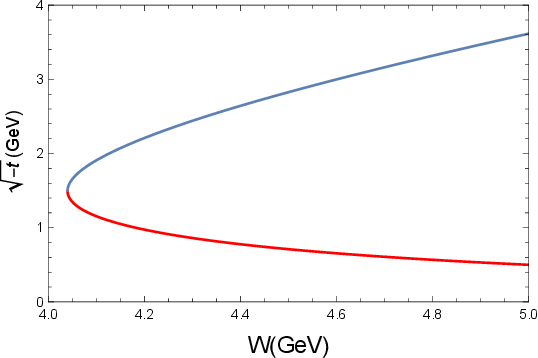}
	\caption{\label{figure}The variation of $\sqrt{-t_{min}}$ (red curve) and $\sqrt{-t_{max}}$ (blue curve) with W for $q_\gamma^2=0$, $M=0.94\ \mathrm{GeV}$ and $M_{V}=M_{J/\psi}=3.10\ \mathrm{GeV}$. Note that $t_{min}=t_{max}$ at the threshold energy $W_{tr}=M+M_{J/\psi}=4.04\ \mathrm{GeV}$.}
\end{figure}

\vspace{1em}

\noindent\textbf{Cross section computation via holography}

\vspace{1em}
In this section, we use holography to calculate the cross section \cite{Hatta:2018ina,Hatta:2010kt,Kruczenski:2003be}. We incorporate charm quarks into the theory through the addition of a D7-brane with the action
 \begin{equation}
\begin{split}
	\label{eq36}
S_{D7} &=-T_{D7}\int d^{8}\xi e^{-\phi}\sqrt{-det(G_{ab}+2\pi\alpha'\digamma_{ab})}\\
 &=-T_{D7}\int d^{8}\xi e^{-\phi}\sqrt{-G}(1+\frac{(2\pi \alpha')^2}{4}\digamma_{ab}\digamma^{ab}+...)
\end{split}
\end{equation}
where $T_{D7}=(32\pi^6g^2\alpha'^4)^{-1}$, $\xi$, $\phi$, $G_{ab}$ and $\digamma_{ab}$ are the $D7$-brane tension, the world-volume coordinates on the $D7$ branes, the dilaton, the induced metric and the gauge field strength tensor respectively. The induced metric can be expressed as
 \begin{equation}
	\label{eq37}
ds_{D7}^2=\frac{R^2}{z^2}\eta_{\mu \nu}dx^\mu dx^\nu-\frac{R^2}{z^2(1-\frac{z^2}{z_m^2})}dz^2-(1-\frac{z^2}{z_m^2})R^2d\Omega_3^2,
\end{equation}
where $R$, $z$ and $z_m$ are the $AdS$ radius, the holographic coordinate and hard truncation, respectively.

$\digamma$ can be expressed as
 \begin{equation}
	\label{eq38}
\digamma=F+\bar{F},
\end{equation}
where $F$ and $\bar{F}$ are the electromagnetic gauge field strength tensors and heavy vector meson field strength tensors, respectively. The wave function of the shell photon can be written as
 \begin{equation}
	\label{eq39}
A_{\mu}\propto \epsilon_{\mu}\chi(z)e^{iq_{\gamma}\cdot x},
\end{equation}
where $\epsilon_{\mu}$ represents the polarization vector, satisfying $\epsilon_{\mu}\cdot q_{\gamma}=0$. The wave function of mesons can be expressed as
 \begin{equation}
	\label{eq40}
\bar{A}_{\mu}\propto \varsigma_{\mu}\varphi(z)e^{ik\cdot x}Y^{l}(S^3),
\end{equation}
where $\varsigma_{\mu}$ represents the vector-meson polarization vector with $\varsigma_{\mu}\cdot k=0$ and $Y^{l}$ represents the spherical harmonics on $S^3$. The mass spectrum can be written as
 \begin{equation}
	\label{eq41}
M_{n,l}=\frac{2\sqrt{(n+l+1)(n+l+2)}}{z_m}.
\end{equation}
We can fix the truncation of $z_m=\frac{2\sqrt{2}}{M_\psi}$ through the ground state mass of $J/\psi$ meson.

We are now starting to calculate the matrix elements $\langle P|\epsilon\cdot J|P'k\rangle$ in the gauge gravity duality. The boundary current insertion $J(q)$ generates a gauge field excitation in the AdS bulk, which subsequently scatters with the proton field through graviton and dilaton exchange processes. This interaction amplitude can be represented as
 \begin{equation}
 \begin{split}
   \label{eq42}
&\langle P|\epsilon\cdot J(q)|P'k\rangle\\
=&\frac{i}{f_\psi}\int d^4xdze^{i(q-k)x}\int d^4x'dz'e^{i(P-P')x'}\\
& [\frac{\delta S_{D7}(q,k,z)}{\delta g_{MN}}G_{MNM'N'}(xzx'z')\frac{\delta S_{B}(P,P',z)}{\delta g_{M'N'}} \\
&+\frac{\delta S_{D7}(q,k,z)}{\delta \phi(xz)}D(xz,x'z')\frac{\delta S_{B}(P,P',z)}{\delta \phi(x'z')}]
\end{split}
\end{equation}
where $f_{\psi}$, $G_{MNM'N'}$, $D$ and $S_{B}$ are the decay constant, the graviton bulk-to-bulk propagator, dilaton bulk-to bulk propagator and the action of baryon field, respectively. This matrix element can be represented by boundary operators $T^{\mu\nu}$ and $F^{\mu\nu}F_{\mu\nu}$ \cite{Hatta:2018ina,deHaro:2000vlm}
\begin{equation}
 \begin{split}
	\label{eq43}
&\langle P|\epsilon\cdot J(q)|P'k\rangle\\
\approx &\frac{-2k^2}{f_\psi R^3}\int_{0}^{z_m}dz \frac{\delta S_{D7}(q,k,z)}{\delta g_{\mu\nu}}\frac{z^2}{R^2} \langle P|T_{\mu\nu}^{gTT}|P'\rangle \\
& +\frac{6k^2}{8f_\psi R^3}\int_{0}^{z_m}dz \frac{\delta S_{D7}(q,k,z)}{\delta \phi}\frac{z^4}{4} \langle P|\frac{1}{4}F^{\mu\nu}_a F_{\mu\nu}^a|P'\rangle,
\end{split}
\end{equation}
where $k=\frac{4\pi^2}{N_c^2}R^3$ and $T_{\mu\nu}^{gTT}$ are the five-dimensional gravitational constant and the transverse-traceless part of the gluon energy-momentum tensor. Notably, only the gluonic $T^{\mu\nu}$ contributes here. This reflects the fact that the graviton is emitted from a $J/\psi$ state, and in QCD heavy quarkonia exclusively couple to gluonic fields rather than light quarks. The holographic realization of this feature stems from the quarkonium-graviton coupling being localized in the asymptotic $AdS$ region ($z\sim 0$), which corresponds to pure gluodynamics with heavy quarks in the dual picture, whereas light quark dynamics are encoded at larger $z$ regions.

From Eq.(\ref{eq36}), we can directly obtain
\begin{equation}
    \label{eq44}
    \begin{split}
        \frac{\delta S_{D7}(q,k,z)}{\delta g_{\mu\nu}}
        = -2K_{D7} &\int d\Omega_3^2 \, Y^{l}(S^3) \frac{z}{R} \varphi(z)\left( 1 - \frac{z^2}{z_m^2} \right) \\
        & \times \left( \Pi^{\mu\nu} - \eta^{\mu\nu} \frac{\eta_{\alpha\beta} \Pi^{\alpha\beta}}{4} \right),
    \end{split}
\end{equation}
 \begin{equation}
     \begin{split}
	\label{eq45}
\frac{\delta S_{D7}(q,k,z)}{\delta \phi}= -K_{D7}&\int d\Omega_3^2 Y^{l}(S^3)\frac{R}{z}\varphi(z)(1-\frac{z^2}{z_m^2})\\
& \times (1+2\frac{z^2}{z_m^2})\frac{\eta_{\alpha\beta}\Pi^{\alpha\beta}}{6},
    \end{split}
\end{equation}
with
 \begin{equation}
	\label{eq46}
\Pi^{\mu\nu}(q_{\gamma},k)=q_{\gamma}^{(\mu} k^{\nu)} \epsilon\cdot \varsigma+\epsilon^{(\mu} \varsigma^{\nu)} q_{\gamma}\cdot k+q_{\gamma}^{(\mu} \varsigma^{\nu)} k\cdot \epsilon+k^{(\mu} \epsilon^{\nu)} q_{\gamma}\cdot \varsigma.
\end{equation}

In actual experimental conditions, direct measurement of the forward matrix element is kinematically forbidden. Instead, experimentalists determine the off-forward matrix element and perform an extrapolation to the forward limit. The general parameterization scheme for the proton's off-forward matrix elements is \cite{Ji:1996ek,Lorce:2018egm}
 \begin{equation}
	\label{eq47}
 \begin{split}
\langle P'|T^{\mu\nu}_{g}|P\rangle &=\bar{u}(P')[(A_{q,g}+B_{q,g})\gamma^{(\mu}\bar{P}^{\nu)}-\frac{\bar{P}^{\mu}\bar{P}^{\nu}}{M}B_{q,g}\\
&+C_{q,g}\frac{\Delta^{\mu}\Delta^{\nu}-\eta^{\mu\nu}\Delta^2}{M}+\bar{C}_{q,g}M\eta^{\mu\nu}]u(P),
 \end{split}
\end{equation}
where $\bar{P}^{\nu}=\frac{P^{\nu}+P'^{\nu}}{2}$ and  $\Delta^\mu=P^\mu-P'^\mu$. $A$, $B$, $C$, and $\bar{C}$ are all gravitational form factors which depend only on $t=\Delta^2$.

Taking the trace of Eq.(\ref{eq47}), we can obtain
\begin{equation}
	\label{eq48}
 \begin{split}
\langle P'|T^{\mu\nu}_{g}|P\rangle & =\langle P'|\frac{\beta(g)}{2g}F^{\mu\nu}_{a}F_{\mu\nu}^{a}+m\gamma_{m}\bar{\Psi}\Psi|P\rangle\\
&=\bar{u}(P')[A_{g}M+\frac{B_{g}\Delta^2}{4M}-\frac{3C_{g}\Delta^2}{M}\\
&\ \ \ \ \ \ \ \ \ \ +4\bar{C}_{g}M]u(P).
 \end{split}
\end{equation}
Then we can obtain
\begin{equation}
	\label{eq49}
 \begin{split}
&\langle P'|\frac{\beta(g)}{2g}F^{\mu\nu}_{a}F_{\mu\nu}^{a}|P\rangle\\ 
=&\bar{u}(P')[(A_{g}-\gamma_{m}A_{q})M+\frac{(B_{g}-\gamma_{m}B_{q})\Delta^2}{4M}\\
&-\frac{3(C_{g}-\gamma_{m}C_{q})\Delta^2}{M}+4(\bar{C}_{g}-\gamma_{m}\bar{C}_{q})M]u(P).
 \end{split}
\end{equation}

The transverse traceless part of $T^{\mu\nu}_{g}$ can be represented as
\begin{equation}
	\label{eq50}
 \begin{split}
\langle P'|T^{\mu\nu}_{gTT}|P\rangle &=\bar{u}(P')[(A_{g}+B_{g})\gamma^{(\mu}\bar{P}^{\nu)}-\frac{\bar{P}^{\mu}\bar{P}^{\nu}}{M}B_{g}\\
&+\frac{1}{3}(\frac{\Delta^{\mu}\Delta^{\nu}}{\Delta^2}-\eta^{\mu\nu})(A_{g}M+\frac{B_{g}\Delta^2}{4M})]u(P).
 \end{split}
\end{equation}

Based on the above discussion, the final expression of the matrix element is
 \begin{equation}
	\label{eq51}
\langle P|\epsilon\cdot J|P'k\rangle=\bar{u}(P')[X\Pi^{\mu\nu}\Gamma_{\mu\nu}+Y\Pi_{\mu}^{\mu}\Gamma]u(P),
\end{equation}
with
\begin{equation}
	\label{eq52}
 \begin{split}
\Gamma_{\mu\nu} &=(A_{g}+B_{g})\gamma^{(\mu}\bar{P}^{\nu)}-\frac{\bar{P}^{\mu}\bar{P}^{\nu}}{M}B_{g}\\
&+\frac{1}{3}(\frac{\Delta^{\mu}\Delta^{\nu}}{\Delta^2}-\eta^{\mu\nu})(A_{g}M+\frac{B_{g}\Delta^2}{4M}),
 \end{split}
\end{equation}
\begin{equation}
	\label{eq53}
 \begin{split}
\Gamma &=\frac{g}{2\beta(g)}((A_{g}-\gamma_{m}A_{q})M+\frac{(B_{g}-\gamma_{m}B_{q})\Delta^2}{4M}\\
&-\frac{3(C_{g}-\gamma_{m}C_{q})\Delta^2}{M}+4(\bar{C}_{g}-\gamma_{m}\bar{C}_{q})M),
 \end{split}
\end{equation}
\begin{equation}
	\label{eq54}
X=\frac{8k^2K_{D7}}{3R^2}\int d\Omega_{3^2}Y^{l=0}(S^3)\int_{0}^{z_{m}}z^3\phi(z)(1-\frac{z^2}{z_m^2})dz,
\end{equation}
\begin{equation}
    \label{eq55}
    \begin{split}
        Y = &\frac{k^2 K_{D7}}{6R^2} \int d\Omega_{3^2} \, Y^{l=0}(S^3)\\ 
        &\times  \int_{0}^{z_{m}} z^3 \phi(z) \left( 1 - \frac{z^2}{z_m^2} \right)\left( 1 + 2\frac{z^2}{z_m^2} \right) \, dz.
    \end{split}
\end{equation}

The first summation encompasses all possible polarization states of both the photon and $J/\psi$ meson, which can be systematically evaluated using the standard polarization sum formula:
\begin{equation}
	\label{eq56}
\sum_{s=1,2}\epsilon_{s}^{\mu}\epsilon^{*\nu}_{s}=-\eta^{\mu\nu}, \sum_{s'=1,2,3}\varsigma_{s'}^{\mu}\varsigma_{s'}^{*\nu}=-\eta^{\mu\nu}+\frac{k^\mu k^\nu}{M_{\psi}^2}.
\end{equation}
The second summation encompasses the spin states of both the initial and final protons. We can obtain
\begin{equation}
	\label{eq57}
 \begin{split}
I_{s} & \equiv \sum_{pol}\sum_{spin}|\langle P|\epsilon\cdot J|P'k\rangle|^2 \\
&=Tr[(X\Pi^{\mu\nu}_{\alpha\beta}\Gamma_{\mu\nu}+Y(\Pi_{\mu}^{\mu})_{\alpha\beta}\Gamma)(\slashed{P}+M)\\
&\ \ \ \ \ \ \ \ (X\Pi^{\mu'\nu',\alpha\beta}\Gamma_{\mu'\nu'}+Y(\Pi_{\mu'}^{\mu',\alpha\beta}\Gamma)(\slashed{P'}+M)]\\
&\ \ \ -\frac{k^\beta k^\gamma}{M_{\psi}^2}Tr[(X\Pi^{\mu\nu}_{\alpha\beta}\Gamma_{\mu\nu}+Y(\Pi_{\mu}^{\mu})_{\alpha\beta}\Gamma)(\slashed{P}+M)\\
&\ \ \ \ \ \ \ \ \ \ \ \ \ (X\Pi^{\mu'\nu',\alpha}_{\ \ \ \ \ \gamma}\Gamma_{\mu'\nu'}+Y(\Pi_{\mu'}^{\mu'})^{\alpha}_{\gamma}) \Gamma(\slashed{P'}+M)]
 \end{split}
\end{equation}
where $\Pi^{\mu\nu}=\Pi^{\mu\nu}_{\alpha\beta}\epsilon^{\alpha}\varsigma^{\beta}$. This expression can be expressed as a function of W and t.

\vspace{1em}

\noindent\textbf{Results}

\vspace{1em}
In this section, we numerically calculate the scattering cross-section. First, since the GFF $C(t)=D(t)/4$ originate from higher-order interactions, we assume $A(t)/D(t)\propto t$ at large $t$. Then the gravitational form factor $D(t)$ can be expressed as
\begin{equation}
	\label{eq58}
D(t)=\frac{-3.6}{1-4.5t}A(t).
\end{equation}
The form of this equation follows our previous work \cite{Deng:2025azp}. The coefficient 3.6 is the average value of $D(0)$ from lattice QCD and other methods, and the coefficient 4.5 is then adjusted to reproduce the lattice QCD results. The ratio $A(t)/D(t)\propto t$ at large $t$ is consistent with the perturbative theory \cite{Tong:2022zax,Broniowski:2025ctl}. Then, by comparing Eq.(\ref{eq27}), Eq.(\ref{eq28}), Eq.(\ref{eq48}) and Eq.(\ref{eq49}), we can obtain
\begin{equation}
	\label{eq61}
b=(A_{q}(0)+4\bar{C}_{q}(0))(1+\gamma_{m}).
\end{equation}
Combining the asymptotic behavior at small t and large t, we assume that the form of $\bar{C}_{g/q}$ is
\begin{equation}
	\label{eq62}
\bar{C}_{g}(t)=-\bar{C}_{q}(t)=\frac{\frac{1-b+\gamma_{m}}{1+\gamma_{m}}-A_{g}(0)}{4(1-\frac{t}{c^2})^2},
\end{equation}
where $c$ is a free parameter.

For any physical process with a transfer momentum of $t$, the effective coupling constant $\alpha_{s}(t)$ can be written as \cite{Tanaka:2018nae}
\begin{equation}
	\label{eq63}
\alpha_{s}(t)=\frac{4\pi}{\beta_{0}\mathrm{log}(\frac{-t}{\Lambda^2})}[1-\frac{\beta_{1}}{\beta_{0}^{2}}\frac{\mathrm{log}(\mathrm{log}\frac{-t}{\Lambda^2})}{\mathrm{log}(\frac{-t}{\Lambda^2})}],
\end{equation}
where $\Lambda=200$\ MeV, $\beta_{0}=\frac{11}{3}N_{c}-\frac{2}{3}N_{f}$ and $\beta_{1}=\frac{34}{3}N_{c}^2-2C_{F}N_{f}-\frac{10}{3}N_{c}N_{f}$ with $C_{F}=\frac{N_{c}^2-1}{2N_{c}}$ represent the leading and sub-leading coefficients of the beta function, respectively. The beta function is \cite{Hatta:2018sqd}
\begin{equation}
	\label{eq64}
\beta(t)=(-\beta_{0}\frac{\alpha_{s}(t)}{4\pi}-\beta_{1}(\frac{\alpha_{s}(t)}{4\pi})^2)g(t),
\end{equation}
where $g(t)=\sqrt{4\pi \alpha_{s}(t)}$. The mass anomalous dimension $\gamma_{m}$ can be represented as
\begin{equation}
	\label{eq65}
\gamma_{m}(t)=6C_{F}\frac{\alpha_{s}(t)}{4\pi}+(6C_{F}^2+\frac{97}{3}C_{F}N_{c}-\frac{10}{3}C_{F}N_{f})(\frac{\alpha_{s}(t)}{4\pi})^2.
\end{equation}
\begin{figure}
	\centering
	\includegraphics[width=8.5cm]{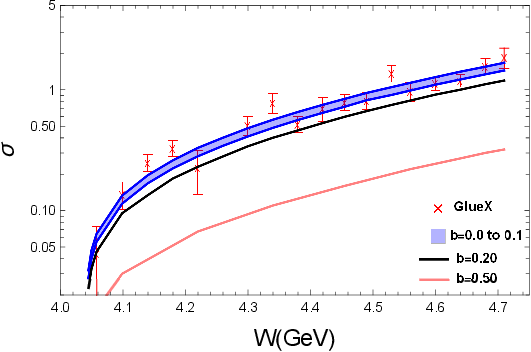}
	\caption{Comparison of the total cross section $\sigma$($\gamma p\rightarrow J/\psi p$) with GlueX data \cite{JointPhysicsAnalysisCenter:2023qgg} for various trace anomaly parameters $b$.}
\end{figure}

As shown in Fig. 6, the total cross-section we calculated is consistent with the experimental results. The calculation involves two free parameters, $b$ and $c$. We choose the most suitable set of parameters based on a $\chi^2$ analysis using the GlueX data: $b = 0.04$. The reduced chi-squared values for different $b$ values are $\chi^2/\text{dof} = 1.175$ (b=0), 1.09 (b=0.04), 1.44 (b=0.10), and 3.43 (b=0.20), indicating that $b=0.04$ gives the best fit. The allowed range from this quantitative analysis is approximately $0 \le b \le 0.10$, which is shaded in Fig. 6, where we adopt the criterion $\chi^{2}/dof<1.5$ for an acceptable fit. Substituting $b = 0.04$ into Eq. (42) yields that the current quarks and the trace anomaly contribute approximately 1.0\% and ~24.0\% (with an allowed range of 22.5–25.0\% given the current experimental uncertainties), respectively, to the proton's mass. This result is consistent with the conclusion of references \cite{Wang:2019mza,Yang:2017erf,Lorce:2017xzd}. The parameter $c=\infty$ implies that no extra phenomenological $t$-dependence is added to $\bar{C}$, and the only $t$-dependence comes from the anomalous dimension $\gamma_m(t)$. Combining Eq.(\ref{eq62}), the gravitational form factor $\bar{C}$ depends on the trace anomaly parameter $b$ and the mass anomaly dimension $\gamma_{m}(t)$, which means it is mainly dominated by gluon condensation and quark condensation. The evolution of the gravitational form factor $\bar{C}(t)$ is reflected through the evolution of the mass anomaly dimension $\gamma_{m}(t)$.

We have also performed a sensitivity check by varying the coefficients in Eq. (70) by $±10\%$ and examining the resulting change in the cross section at the trace anomaly parameter $b=0.04$. Over the entire energy range of our calculation, the cross section changes by less than $1\%$ under this variation, indicating that the extracted value of $b$ is stable against the uncertainty in these coefficients.

As can be further observed from Fig. 6, the larger the parameter $b$ (corresponding to a smaller trace anomaly), the smaller the total cross section. In Regge theory, the high-energy scattering $\gamma p\rightarrow J/\psi p$ is dominated by Pomeron exchange, whose coupling strength is intrinsically linked to non-perturbative properties of the gluon field (e.g., gluon condensation). The observed reduction in the trace anomaly signifies a suppression of gluon field non-perturbative effects, thereby diminishing the Pomeron coupling strength and consequently decreasing the total scattering cross section.
\begin{figure}
	\centering
	\includegraphics[width=8.5cm]{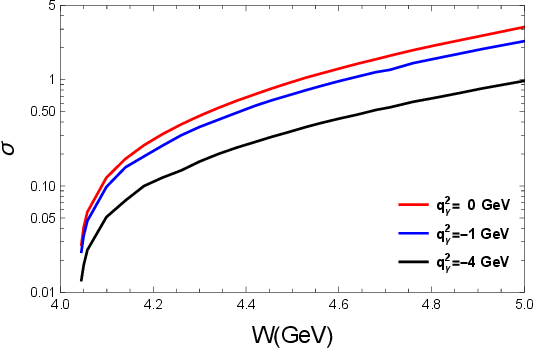}
	\caption{\label{figure}Total cross section $\sigma$ for different values of the photon virtuality $q_\gamma^2$.}
\end{figure}

Finally, we plot the total cross-section $\sigma$ for different $q_\gamma^2$ values in Fig.7. $q_\gamma^2=0$ and $q_\gamma^2<0$ correspond to real photons and virtual photons, respectively. We can see that the total cross-section corresponding to virtual photons is smaller. The trend is consistent with calculations from other model \cite{Mamo:2021tzd}. This is because as $q_\gamma^2$ increases, the decay rate of photons becomes faster, resulting in a decrease in the interaction with quark gluons inside protons.

\section{Conclusion}\label{sec:04}

In this study, we have constructed the quark and gluon GPDs within a phenomenological extension of light-front holographic QCD (LFHQCD), starting from the electromagnetic form factors provided by LFHQCD. From these GPDs, we then derive the corresponding parton distribution functions (PDFs) and gravitational form factors (GFFs) of the proton. Our approach, which incorporates a parametrized function respecting both the inclusive counting rules at large momentum fraction $x$ and Regge theory constraints at small $x$ , has yielded results in good agreement with available experimental data and lattice QCD calculations. This consistency supports the effectiveness of the LFHQCD framework and our phenomenological extension in describing the non-perturbative structure of the proton.

The central role of the Pomeron in our analysis of gluonic contributions merits further discussion. In QCD, the Pomeron, crucial for small-angle scattering, is interpreted as a $2^{++}$ two-gluon bound state. Its dual description in AdS space corresponds to a graviton. By assuming that the massless graviton ground state decouples from strong interactions and that its first excited state manifests as the $2^{++}$ glueball, we obtained a glueball mass in the range $M = (2.0 - 2.24)$ GeV. This prediction aligns remarkably well with results from lattice QCD simulations and other theoretical models, providing strong support for the holographic interpretation of Pomeron physics and its connection to glueball states.

A key achievement of this work is the quantitative evaluation of the QCD trace anomaly's contribution to the proton mass. By calculating the $J/\psi$ photoproduction cross section using our complete set of GFFs and incorporating the energy-scale dependence of the running coupling constant (via the two-loop QCD $\beta$-function) and the anomalous dimensions $\gamma_m$, we found excellent agreement with experimental measurements. This analysis yielded a trace anomaly contribution of approximately 24.0\% to the proton mass. This result is significant as it bridges insights from different methodologies: it is consistent with both direct experimental extractions and contemporary lattice QCD predictions, reinforcing the understanding that a substantial fraction of the proton's mass originates not from the Higgs mechanism but from the dynamical effects of QCD, with the trace anomaly being a key manifestation.

While our results show compelling agreement with existing data, several limitations should be acknowledged. Firstly, in constructing the gluon GPDs, we considered only the lowest Fock state contribution. Higher Fock states and more complex gluon configurations could introduce systematic uncertainties, particularly in the small-$x$ region where gluon densities are high and nonlinear effects may become important. Secondly, our parametrization, while constrained by general principles, represents a specific model choice. Exploring alternative functional forms that satisfy the same boundary conditions could help quantify model-dependent uncertainties. Finally, while the present framework successfully describes the energy dependence of the total cross section $\sigma(W)$, its application to the differential cross section $d\sigma/dt$ reveals limitations. This shortcoming stems from the perturbative treatment of the running coupling, which is insufficient for describing the detailed momentum transfer dependence governed by non-perturbative QCD.

These limitations naturally point to directions for future research. Subsequent studies should aim to incorporate a more comprehensive treatment of gluon dynamics, potentially by including higher-twist effects and exploring the impact of gluon saturation at very small $x$. Furthermore, applying this holographic framework to other exclusive processes, such as deeply virtual Compton scattering (DVCS) or timelike Compton scattering, would provide additional stringent tests of the derived GPDs. Extending the analysis to other hadrons, like neutrons or mesons, could offer deeper insights into the universality of the mechanisms discussed here. Finally, a more detailed microscopic study of the glueball spectrum and its coupling within the holographic model could further elucidate the connection between Pomeron exchange and gluonic confinement dynamics. Future work should also incorporate a non-perturbative running coupling to enable comparison with $d\sigma/dt$ data from GlueX.

In conclusion, this work demonstrates the power of combining LFHQCD with constraints from Regge behavior and counting rules to build a consistent picture of the proton's quark and gluon structure. The successful description of GPDs, PDFs, and GFFs, along with the calculated $J/\psi$ cross sections, provides robust evidence for a significant ($\sim24.0\%$) trace anomaly contribution to the proton mass. Our findings not only align with independent theoretical and experimental results but also solidify the holographic approach as a valuable tool for probing the non-perturbative regime of QCD.

\section*{Acknowledgments}

We thank Hai-cang Ren, Craig D. Roberts, Minghui Ding and  Yan-Qing Zhao for useful discussions. This work is supported in part by the National Key Research and Development Program of China under Contract No. 2022YFA1604900. This work is also partly supported by the National Natural Science Foundation of China(NSFC) under Grants No. 12435009, and No. 12275104.

\section*{References}

\bibliography{ref}
\end{document}